\documentclass[twoside,a4paper,11pt]{proceedings}
\usepackage{graphicx}
\usepackage{hyperref}
\usepackage{movie15}
\usepackage{natbib}
\usepackage{wrapfig}

\begin{document}
\pagenumbering{arabic}
\pagestyle{myheadings}
\thispagestyle{empty}
\vspace*{-1cm}
{\flushleft\includegraphics[width=3cm,viewport=0 -30 200 -20]{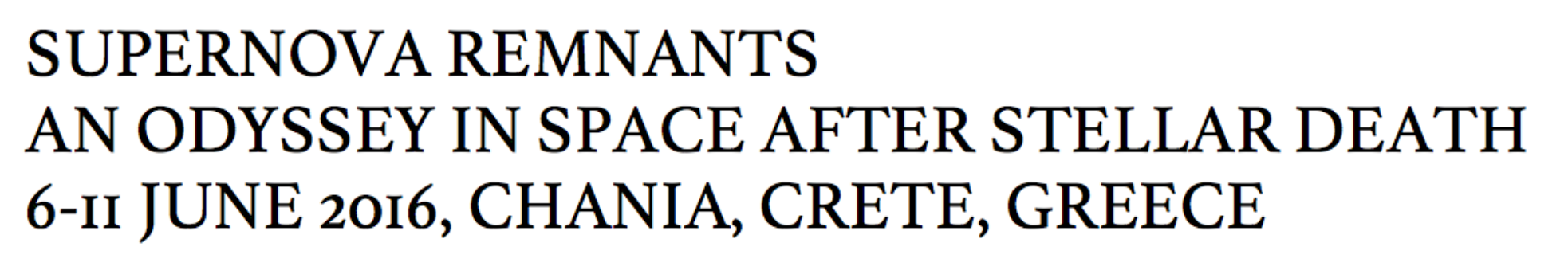}}
\vspace*{0.2cm}
\begin{flushleft}
{\bf {\LARGE
Dust grains from the heart of supernovae
}\\
\vspace*{1cm}
M. Bocchio$^{1,2}$,
S. Marassi$^2$,
R. Schneider$^2$
S. Bianchi$^1$
M. Limongi$^2$
A. Chieffi$^3$
%
}\\
\vspace*{0.5cm}
%
$^{1}$
INAF - Osservatorio Astrofisico di Arcetri, Largo Enrico Fermi 5, 50125 Firenze, Italy \\
$^{2}$
INAF - Osservatorio Astronomico di Roma, Via di Frascati 33, I-00040 Monteporzio, Italy \\
$^{3}$
INAF/IASF, Via Fosso del Cavaliere 100, 00133 Roma, Italy \\
%
\end{flushleft}
\markboth{
Dust grains from the heart of supernovae
}{
M. Bocchio et al.
}
\thispagestyle{empty}
\vspace*{0.4cm}
\begin{minipage}[l]{0.09\textwidth}
\ 
\end{minipage}
\begin{minipage}[r]{0.9\textwidth}
\vspace{.5cm}
\section*{Abstract}{\small
Dust grains are classically thought to form in the winds of asymptotic giant branch (AGB) stars. However, there is
increasing evidence today for dust formation in supernovae (SNe). To establish the relative importance of these two
classes of stellar sources of dust, it is important to know the fraction of freshly formed dust in SN ejecta that is able to
survive the passage of the reverse shock and be injected in the interstellar medium. 

We have developed a new code (GRASH\_Rev) which follows the
newly-formed dust evolution throughout the supernova explosion until the merging of the forward
shock with the circumstellar ISM. 

We have considered four well studied SNe in the Milky Way and Large Magellanic Cloud:
SN1987A, CasA, the Crab Nebula, and N49. For all the simulated models, we find good agreement with observations
and estimate that between 1 and 8\% of the observed mass will survive, leading to a SN dust production rate of 
$(3.9 \pm 3.7) \times 10^{-4}$\,M$_{\odot}$yr$^{-1}$ in the Milky Way. This value is one order of magnitude larger than the dust production
rate by AGB stars but insufficient to counterbalance the dust destruction by SNe, therefore requiring dust accretion 
in the gas phase.

\vspace{5mm}
\normalsize}
\end{minipage}

\section{Introduction}

It is observationally and theoretically well established that a considerable amount of dust is efficiently formed in regions around asymptotic giant branch (AGB) stars. This process is classically
considered as the primary source of dust grains in galaxies, and the typical formation timescale in
the Milky Way is $\sim 3\times 10^9$\,yr.

In contrast, supernova (SN) explosions in the interstellar medium (ISM) trigger shock waves
that are able to quickly process dust grains and are considered the dominant mechanism of dust
destruction in the ISM. Recent theoretical and observational work on interstellar dust destruction in shock waves
led to an estimated dust lifetime much shorter than the assumed dust
formation timescale from AGB stars. 

Although SNe are believed to be efficient interstellar dust destroyers, there is increasing observational 
evidence today for the formation of non-negligible quantities of dust grains in the ejecta
of SNe. Given the relatively short timescale between the explosion of two SNe, this would lead to
an effectively shorter timescale for dust formation.

In this work we present a new code called GRASH\_Rev that
treats dust processing in a supernova explosion. This code couples
all the dust processing included in the GRASH\_EX code
(Bocchio et al. 2014) with the dynamics and structure of the SN
as modelled by Bianchi \& Schneider (2007, BS07) but extending it to include the full dynamics
of dust grains within the ejecta and in the surrounding ISM.

\vspace{-.2cm}
\section{Supernova sample}

Only in four core-collapse SNe it has been possible to
estimate the amount of dust in the ejecta, using observation
from Spitzer and Herschel: SN 1987A, Cassiopeia A,
Crab Nebula and SNR N49.
We obtain their main physical properties from the literature:
the type of SN, the explosion energy ($E_{\rm ex}$),
the mass of $^{56}$Ni, the progenitor mass ($M_{\rm prog}$), the estimated
metallicity ($Z$) of the progenitor star, the age, the number density
of the circumstellar ISM ($n_0$) and the measured dust mass associated
with the ejecta ($M_{\rm dust}$).

A comparison between the dust mass observed in these objects and the
amount and composition of dust that our model predicts gives constraints on the
ongoing physical processes.
This makes our model reliable and allows us to predict the amount of dust that
will be released into the ISM.
\vspace{-.2cm}
\section{The GRASH\_Rev code}

Starting from a homogeneous set of solar metallicity pre-supernova
models with masses in the range [13-120] $M_{\odot}$
(Chieffi \& Limongi 2013) simulated by means of the FRANEC
stellar evolutionary code (Limongi \& Chieffi 2006), we have selected
the most suitable model for each SN according to their
physical properties. These four selected models
are then used as input for the dust formation code (BS07),
where classical nucleation theory in steady state conditions was
applied. Here we use the latest version of dust formation model,
which implements an upgraded molecular network, as described
in detail in Marassi et al. (2015). We follow the formation of
six different dust grain species: amorphous carbon (AC), corundum
(Al2O3), magnetite (Fe3O4), enstatite (MgSiO3), forsterite
(Mg2SiO4) and quartz (SiO2). 

We then adopt the analytical solution described by Truelove \& McKee (1999) and
Cioffi et al. (1988) to characterise the SNR evolution from the early ejecta-dominated phase through
the pressure-driven snowplough phase until the velocity of the shock front becomes low
enough ($\sim$ 10 km/s) and the remnant merges with the ISM.

The dynamics of dust is entirely governed by its coupling to the gas,
which in the GRASH\_Rev code is described as the combination of the collisional (direct)
and plasma drag.
We then consider four main physical processes: sputtering due to the interaction of dust grains
with particles in the gas; sublimation due to collisional heating to high temperatures;
shattering into smaller grains due to grain-grain collisions and 
vapourisation of part of the colliding grains during grain-grain collisions.

\begin{wrapfigure}{r}{0.47\textwidth}
\vspace{-35pt}
  \begin{center}
    \includegraphics[width=0.45\textwidth, clip=1,trim=1cm 2cm 1cm 4cm]{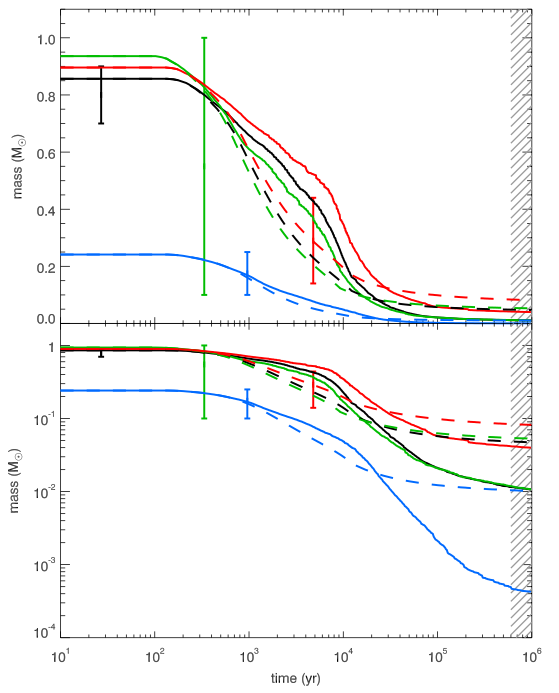}
  \end{center}
  \caption{\small Dust mass evolution as a function of time (1987A: black, CasA: green, Crab:
blue and N49: red). Solid and dashed lines are GRASH\_Rev and BS07 results. 
Data points are observed dust masses and the shaded region indicates the time interval when
dust processing fades out.}
\vspace{-35pt}
\end{wrapfigure}
\vspace{-.2cm}

\section{Results}

We ran simulations for the four SNe considered and follow the dust mass
evolution. 
In Fig. 1 we show the time evolution of the dust mass following
the SN explosion. 
The dust mass estimated using GRASH\_Rev simulations are
in good agreement with the observations, except for SNR N49
for which the dust mass predicted by GRASH Rev is $\sim15\%$
larger than the upper mass limit of the range of values inferred
from observations.
The resulting dust yields depend on the progenitor
mass, explosion energy and the age of the SN, and lie in the
range $4\times10^{-4} - 4\times10^{-2}$\,M$_{\odot}$.

\vspace{-.2cm}

\section{Discussion and conclusions}

Given the rather large uncertainties on measurements of dust
masses, our models appear to reproduce the dust masses in
the four SNe well. The average effective dust yield is estimated
to be (1.55 $\pm$ 1.48) 10$^{-2}$ M$_{\odot}$.
When compared to dust destruction efficiencies in SN-driven
interstellar shocks that were recently estimated by theoretical
models (Bocchio et al. 2014; Slavin et al. 2015) and
observations (Laki\'cevi\'c et al. 2015), this implies that SNe may
be net dust destroyers, pointing to grain growth in the ISM as
the dominant dust enrichment process both in local galaxies
and at high redshifts.

\vspace{-.2cm}
\small  
%
\section*{Acknowledgments}   
%
{\small The research leading to these results has received funding
from the European Research Council under the European Union (FP/2007-2013)
/ ERC Grant Agreement n. 306476.}
\vspace{-.2cm}
\section*{References}
\bibliographystyle{aj}
\small
\bibliography{proceedings}
Bocchio et al., 2014, A\&A, 570, A32\\
Bianchi \& Schneider 2007, MNRAS, 378, 973\\ 
Chieffi \& Limongi 2013, ApJ, 764\\
Limongi \& Chieffi 2006, ApJ, 647, 483\\
Marassi et al. 2015, MNRAS, 454, 4250-4266\\
Slavin et al. 2015, ApJ, 803, 7\\
Laki\'cevi\'c et al. 2015, ApJ, 799, 50\\
\end{document}